\documentclass[10pt, conference, compsocconf]{IEEEtran}
\usepackage{color}
\usepackage{xcolor}
\usepackage{setspace}

\newcommand{\ie}{{\it i.e.}}

\newcommand{\etc}{{\it etc}}

\newcommand{\RNum}[1]{\uppercase\expandafter{\romannumeral #1\relax}}
\newcommand{\specialcell}[2][c]{%
  \begin{tabular}[#1]{@{}c@{}}#2\end{tabular}}

\usepackage{array}
\newcolumntype{L}[1]{>{\raggedright\let\newline\\\arraybackslash\hspace{0pt}}m{#1}}
\newcolumntype{C}[1]{>{\centering\let\newline\\\arraybackslash\hspace{0pt}}m{#1}}
\newcolumntype{R}[1]{>{\raggedleft\let\newline\\\arraybackslash\hspace{0pt}}m{#1}}
\let\labelindent\relax
\usepackage{enumitem}
\usepackage{textcomp}
\usepackage{graphicx}
\graphicspath{ {images/} }

\begin{document}

\title{A Survey of Semantics-Aware Performance Optimization for Data-Intensive Computing}


\author{
\IEEEauthorblockN{Bingbing Rao}
\IEEEauthorblockA{Department of Computer Science\\
University of Central Florida\\
Orlando,Florida,U.S.\\
Robin.Rao@knights.ucf.edu}
\and
\IEEEauthorblockN{Liqiang Wang}
\IEEEauthorblockA{Department of Computer Science\\
University of Central Florida\\
Orlando,Florida,U.S.\\
lwang@cs.ucf.edu}
}

\maketitle

\begin{abstract}
We are living in the era of Big Data and witnessing the explosion of data. Given that the limitation of CPU and I/O in a single computer, the mainstream approach to scalability is to distribute computations among a large number of processing nodes in a cluster or cloud. This paradigm gives rise to the term of data-intensive computing, which denotes a data parallel approach to process massive volume of data. Through the efforts of different disciplines, several promising programming models and a few platforms have been proposed for data-intensive computing, such as MapReduce, Hadoop, Apache Spark and Dyrad. Even though a large body of research work has being proposed to improve overall performance of these platforms, there is still a gap between the actual performance demand and the capability of current commodity systems. This paper is aimed to provide a comprehensive understanding about current semantics-aware approaches to improve the performance of data-intensive computing. We first introduce common characteristics and paradigm shifts in the evolution of data-intensive computing, as well as contemporary programming models and technologies. We then propose four kinds of performance defects and survey the state-of-the-art semantics-aware techniques. Finally, we discuss the research challenges and opportunities in the field of semantics-aware performance optimization for data-intensive computing.


\end{abstract}


%
\IEEEpeerreviewmaketitle

\section{Introduction}


With data explosion in many areas, such as scientific experiments, telescopes, e-commerce, social media \cite{chung2017itsec} and financial service data, people are facing critical problems on data processing and analytics. 
In many instances, science is falling far behind reality in the capabilities of exploring the valuable knowledge from enormous volume of data. Consequently, we need to investigate and develop novel techniques to excavate data-intensive issues \cite{subramanian2011rapid,subramanian2010rapid}. Jim Gray \cite{hey2009fourth} declared that ``the techniques and technologies for such data-intensive science are so different that it is worth distinguishing data-intensive science from computational science as a new, fourth paradigm for scientific exploration", in contrast to the prior three, empirical science, theoretical science, and computational science. 

Through the efforts of different disciplines, a few promising programming models and platforms have been proposed for data-intensive computing, such as MapReduce \cite{dean2008mapreduce}, Hadoop \cite{hadoop2009hadoop}, Apache Spark \cite{zaharia2010spark}, Dryad \cite{isard2007dryad}, Pregel \cite{malewicz2010pregel}, and Dremel \cite{melnik2010dremel}. However, performance limitation is always an encumbrance of these platforms even they can support processing massive scale of data. In order to improve overall performance of these platforms, researchers and participants have been developing a bunch of optimization techniques via various methodologies from different dimensions, such as improving query processing and data management in a distributed file system, applying machine learning technologies for determining an optimal parameter configuration for systems and reducing data movement, enhancing memory management to minimize the amount of data shuffling between two different stages and lessen the memory pressures based on program analysis techniques.

There is a large body of research work involved in performance optimizations based on semantics-aware technology.
The semantic information of data has a great effect on data management issues such as data capture and data store; the structure information also affects data partitions. For now, the common algorithms for partitioning are based on hash and range functions. In the world of program analysis, the philosophy behind code guides the behavior of system performance. For example, in a data-driven application, filtering and removing unused data before a shuffling operation is crucial for system performance since this kind of optimization can reduce the amount of data shuffling. Even though it is evident that scaling out (\ie, adding more nodes, {\it a.k.a.}, horizontally) is a fast way to improve system performance, improving performance of each individual node, \ie, scaling up vertically, is another option to improve system performance. Memory management is also a critical issue, especially for in-memory systems. How to optimize data structure and manage objects efficiently is a major challenge that we confront in a data-intensive computing system. The interactions among data, code and system are made up of a comprehensive methodology of data-intensive computing system, which is crucial to the performance of these platforms. Unfortunately, there is still a gap between actual demand and system performance. In this study, we survey semantics-aware approaches to optimize the overall system performance.


The rest of paper is organized as follows. In Section \ref{sec:background}, we introduce the common characteristics and several influential paradigm shifts in the evolution of data-intensive computing. The contemporary programming models and technologies are reviewed in Section \ref{sec:methodology}. Section \ref{sec:performance} concentrates on performance defects of data-intensive computing from the perspective of semantics-aware technology. The current research challenges and opportunities are discussed in Section \ref{sec:research}. Finally, conclusions are presented in Section \ref{sec:conclusion}.

\section{Background: Characteristics and Paradigm Shifts}
\label{sec:background}

\subsection{Common Characteristics}
In comparison with other forms of computing, there are several crucial characteristics of data-intensive computing systems \cite{middleton2011hpcc,wiki:xxx}.
\begin{enumerate}
	\item \textit{Move code to data:} The generation of an enormous amount of data has led to the proliferation of data-intensive applications. In the data-intensive computing ecosystem, distributed file systems \cite{shvachko2010hadoop} split data among machines so that many processing nodes can work simultaneously. In this scenario, a program or algorithm is transferred to nodes where data reside, and avoid moving data to computing nodes as much as possible, which is also know as ``move code to data". Obviously, the key to achieve high performance in data-intensive computing is to minimize the amount of data shuffling between two different stages.
    
    \item \textit{Programming model:} Data-intensive computing systems leverage a machine-independent approach to partition a problem into concurrent tasks, each of which is expressed in terms of high-level operations on data \cite{bryant2007data}. To counterbalance overall performance of applications, the runtime system transparently orchestrates scheduling, execution, load balancing, communications, and movement of programs and data across distributed computing nodes. Therefore, we need programming models and language tools to formalize data flows and transformations in assimilating new dataflow programming languages and sharing libraries of common data manipulation algorithms such as join or sort operations.
    
    \item \textit{Reliability and availability:} Data-intensive computing systems with hundreds or thousands of processing nodes are more sensitive to hardware failures, communications errors and software bugs. In order to enable systems to continue operating properly in the event of failures, data-intensive computing systems should be fault-tolerant. For example, there should be a data replication mechanism in the storage system to recover data when data is missing. The runtime systems have an ability to store intermediate processing results on disk or in memory and monitor status of computing nodes to recover from an incomplete processing automatically and transparently.  
 
    \item \textit{Inherent scalability:} The mainstream approach to scalability is to distribute computation among a large number of processing nodes in clusters or in cloud. Data-Intensive computing systems can be scaled out horizontally for accommodating massive volumes of data in which computations on data can be distributed to processing nodes. 

\end{enumerate}

\subsection{Paradigm Shifts}
With the increasingly expanding of datasets, there is an increasing amount of diversified challenging issues regarding processing massive volume of data for many decades. 
Since data size increases much faster than computing resources. In order to cope with these issues, paradigm shifts rapidly \cite{chen2014data}.

\begin{enumerate}
	
	\item \textit{A shift in processor technology:}
	Even though the clock cycle frequency of processors was doubling approximately  every 18 months following Moore's law, because of power supply constraints, the clock speed highly lags behind. Alternatively, people made an effort to increase numbers of cores in a single processor to make application run in parallel \cite{almasi1988highly, asanovic2006landscape}. 
Unfortunately, with the limitation of hardware technologies, it is very hard to embed as much number of cores in a single processor as we want. Hence, scientists started to initiate data-intensive computing, in which a series of jobs are distributed to different computing nodes containing the needed raw data. Even though data-intensive computing has been proposed and studied for a few years, it is still a big challenge for data processing system to optimize parallelism across nodes in a cluster. 

	\item \textit{A shift in I/O subsystem:}
	Data-intensive computing has been changing the way to capture and store data \cite{oliveira2012trends}, which includes data storage device, architecture, as well as data access mechanism. In the initial, hard disk drive (HDD) with slower random I/O performance is used to store persistent data. In order to make data accessed easily and promptly for further analysis, data processing engine formats data and uses specialized query processing techniques to alleviate this limitation. With the development of storage technologies, solid-state drive (SSD) and phase-change memory (PCM) help mitigate the difficulty, but which are far from enough. There are severe drawbacks and limitations among existing storage architectures when it comes to data-intensive computing systems. For example, how to store and access data efficiently in a distributed file system is still a challenge issue for data-intensive computing frameworks. In consequence, a newer storage subsystem  needs to be redesigned for large-scale distributed systems.


	\item \textit{A shift in scientific investigation:}
	Contemporary scientific research demands new data mining tools, novel mathematical and statistical techniques, advanced machine learning algorithms as well as other data analytical disciplines, to facilitate the process of data-intensive computing problems \cite{chen2014data}. For example, it is very difficult for many traditional methods that perform well on small data to scale up on massive data. 
We should take data-intensive computing problems, such as heterogeneity, noise accumulation, spurious correlations and incidental endogeneity, into consideration to design effective statistical and machine learning methods for harnessing data-intensive computing issues, in addition to balancing the statistical accuracy and computational efficiency \cite{fan2014challenges}.
\end{enumerate}

\section{Methodology: Programming Models and Processing Technologies}
\label{sec:methodology}
With massive data generated from various fields, it is necessary to develop novel programming models and processing technologies to cope with data-intensive computing issues. From the perspective of different disciplines, there is a wide variety of programming and platforms that have been developed to meet their specific purposes. However, they cannot match practical needs. In this section, we survey  contemporary programming models and processing technologies on data-intensive computing systems.

\subsection{Programming Models}
Data-intensive computing needs programming models to handle enormous volume of data efficiently. A programming model represents a style of programming and interface paradigm for developers to encode applications, as well as provides a way to control scheduling, execution, load balancing, communications, and movement of programs and data across distributed computing nodes. Academia and industries have been proposing and developing a set of data-intensive programming models, as illustrated in Table \ref{tab:programming_model}. In this section, we will discuss and compare two major programming models for data-intensive computing, MapReduce and Functional Programming.
\begin{table}[ht]
	\fontsize{6}{7.2}
	\selectfont
	\begin{center}
    	\begin{tabular}{|C{1.5cm}|C{5cm}|C{1cm}|}
        \hline
        	Programming model&
            Features&
            Examples\\
        \end{tabular}
        \begin{tabular}{|C{1.5cm}|C{5cm}|C{1cm}|}
        \hline
        	MapReduce&
            \begin{enumerate}[leftmargin=2mm]
				\item[1.] A simple paradigm with functionalities of Map and Reduce.
				\item[2.] Key-value pairs provides good support of parallelization and scalability.
				\item[3.] Parallelable and scalable to hundreds or thousands of processing nodes.
                \item[4.] Tolerate machine failures gracefully.
			\end{enumerate}&
            \specialcell{MapReduce\\Hadoop}\\
        \end{tabular}
        \begin{tabular}{|C{1.5cm}|C{5cm}|C{1cm}|}
        \hline
        	Functional Programming&
            \begin{enumerate}[leftmargin=2mm]
				\item[1.] Specify semantic logic of computation declaratively.
				\item[2.] A philosophical match between functional programming and parallelism based on immutable feature.
				\item[3.] Use tail-recursive approaches to reduce intermediate data and variables shared in different loops.
				\item[4.] The features of high-order functions and type inference are convenient to implement machine learning algorithms.
                \item[5.] Process the input data as streams format.
                \item[6.] Lambda expressions provides a good way to define data operation functions.
			\end{enumerate}&
            \specialcell{Spark\\Flink}\\
        \end{tabular}
        
        \begin{tabular}{|C{1.5cm}|C{5cm}|C{1cm}|}
        \hline
        	SQL-Based&
            \begin{enumerate}[leftmargin=2mm]
				\item[1.] Support declarative programming paradigm. 
				\item[2.] One of the most popular programming model for data-centric applications using data-driven operations.
				\item[3.] Standard protocol supports interoperability between different platforms and frameworks.
                \end{enumerate}&
            \specialcell{HiveQL\\CasandraQL\\SparkSQL\\Drill}\\
        \end{tabular}
        
        \begin{tabular}{|C{1.5cm}|C{5cm}|C{1cm}|}
        \hline
        	Actor Model&
            \begin{enumerate}[leftmargin=2mm]
            \item[1.] A message-oriented architecture for communicating.
				\item[2.] Stateless and isolation among different actors.
				\item[3.] Support concurrency based on actor mechanisms.
			\end{enumerate}&
            \specialcell{Akka\\Storm\\S4}\\
        \end{tabular}
        
        \begin{tabular}{|C{1.5cm}|C{5cm}|C{1cm}|}
        \hline
        	Statistical and Analytical&
            \begin{enumerate}[leftmargin=2mm]
           \item [1.]Support declarative programming paradigm. 			
            \item[2.] A comprehensive and encapsulated API in function format. 
			\item[3.] Matrix-based data structure in computations.			
			\end{enumerate}&
            \specialcell{R\\Mahout}\\
        \end{tabular}
        \begin{tabular}{|C{1.5cm}|C{5cm}|C{1cm}|}
        \hline
        	Dataflow-Based&
            \begin{enumerate}[leftmargin=2mm]
				\item[1.] Provide a trackable state during execution since programs are treated as connections of tasks in combination with control logic.
				\item[2.] Flexible ways of definition, such as graph-based manner and Hash tables.
			\end{enumerate}&
            \specialcell{Oozie\\Dryad}\\
        \end{tabular}
        
        \begin{tabular}{|C{1.5cm}|C{5cm}|C{1cm}|}
        \hline
        	Bulk Synchronous Parallel&
            \begin{enumerate}[leftmargin=2mm]
				\item[1.] Message-based communication that reduces the effort for users to handle low-level parallel communications.
				\item[2.] A barrel-based synchronization mechanism that guarantees consistency and fault tolerance in an easy and understandable way.
			\end{enumerate}&
            \specialcell{Giraph\\Hama}\\
        \end{tabular}
        
        \begin{tabular}{|C{1.5cm}|C{5cm}|C{1cm}|}
        \hline
        	High-Level DSL&
            \begin{enumerate}[leftmargin=2mm]
				\item[1.] Provide Domain Specific Language (DSL) model to specify data-intensive applications.
		 	\end{enumerate}&
            \specialcell{Pig Latin\\Jaql\\AQL\\LINQ}\\
        \hline
        \end{tabular}
        
    	\caption{Taxonomy of programming models on data-intensive computing}
		\label{tab:programming_model}
    \end{center}
\end{table}

\begin{enumerate}
	\item 
 	\textit{MapReduce} is a framework for programming commodity computer clusters to perform large-scale data processing in a single pass \cite{dean2008mapreduce}. For a single MapReduce job, programmers implement two basic procedure objects, Mapper and Reducer, to present users' logical plan on dataset. The functionality of Mapper object performs filtering and sorting operations on input dataset and generates a series of intermediate data with key-value pair format as the inputs of Reducer objects. Reducer method performs an aggregate operation, such as counting the number of students in each queue and yielding name frequencies. 
Apache Hadoop \cite{hadoop2009hadoop} is an open-source implementation of Google's MapReduce paradigm \cite{dean2008mapreduce}. 
In the Hadoop ecosystem, platform uses a distributed file system, called Hadoop Distributed File System (HDFS) \cite{shvachko2010hadoop}, to provide high throughput access to massive data. 
Fault-tolerance and dynamic scalability support adding or removing computing nodes without altering the existing systems and programs, which makes it one of most widespread systems in the data-intensive ecosystem. 

	\item 
    \textit{Functional Programming} is a style of programming that supports immutable state, higher order functions, type inference, the processing of data as streams, lambda expressions, and concurrency through software transactional memory. Because of these features, it is becoming a novel paradigm for the next generation of data-intensive computing systems \cite{zomaya2017handbook}. Declarative manner in functional programming provides an easier and more convenient way for users to specify the semantic logic of computation, rather than the control flow of procedures. In principle, states in a functional program are immutable, which means that states cannot be modified, \ie, no side effects. The features of high order functions, which define a program in a functional manner and take one or more functions as arguments, are convenient for algorithm design when passing functions as parameters. For example, when designing a machine learning algorithm, it is feasible to pass different regularizes, update rules, or even learning algorithms altogether as function parameters. Type inference system also gives programmers a way to implement algorithms efficiently since it is not necessary to pay special attentions to type information. There is a great deal of platforms adoring functional programming languages, for example, Apache Spark \cite{zaharia2010spark} and Flink \cite{alexandrov2014stratosphere} utilize features of functional language to facilitate developers to design data-intensive applications in an easy and declarative way.

\end{enumerate}

\subsection{Processing Technologies}
It is crucial to explore a series of tools to solve data-intensive issues. To the best of our knowledge, there is a wide variety of classifications based on different dimensions \cite{chen2014data,katal2013big,  khan2014big}. In this paper, we concentrate on the following four classes, \ie, query processing, batch processing, stream processing, and interactive processing.
\begin{enumerate}
	
	\item 
	\textit{Batch processing} is an execution of a bunch of jobs in a program that take a set of data files as input, process the data, and produce a list of output data files. In recent years, there are many batch processing systems proposed, such as MapReduce \cite{dean2008mapreduce}, Hadoop \cite{hadoop2009hadoop}, Spark \cite{zaharia2010spark} and Pregal \cite{malewicz2010pregel}. These systems analyze large dataset in batches in a distributed and parallel fashion. In particular, Apache Spark \cite{zaharia2010spark} is a fast and general engine for large-scale data processing that supports scalability and fault tolerance of MapReduce \cite{dean2008mapreduce}. Apache Spark introduces a distributed memory abstraction, named Resilient Distributed Datasets (RDD) \cite{zaharia2012resilient}, to support in-memory computations across multiple nodes in a fault-tolerant manner.
	
	\item
	\textit{Streaming processing} is a real-time system that processes continuous input of data. In a real-time system, data processing requires fast response, which means the rate of  processing should be not slower than the rate of incoming data. Data-intensive streaming platforms include Storm \cite{toshniwal2014storm}, Spark Streaming \cite{zaharia2012discretized}, S4 \cite{neumeyer2010s4}, \etc. Spark Streaming is an internal component of Apache Spark that enables scalable, high-throughput, fault-tolerant processing of live data streams. Data can be taken from many sources like Kafka, Flume, Kinesis, or TCP sockets, and can be processed using complex algorithms expressed with high-level functions like map, reduce, join and window. Finally, results are output to file systems, databases, or live dashboards.

	\item
	\textit{Query processing} is a platform that can translate user queries into data retrieval and processing operations, and execute these operations on one or multiple nodes. Many distributed computing platforms like Hive \cite{thusoo2009hive}, Pig Latin\cite{olston2008pig}, and Spark SQL \cite{armbrust2015spark} are query processing systems. In these frameworks, programmers use a declarative manner to specify their jobs, which are then translated into appropriate optimized operations. Spark SQL is built atop of Apache Spark to integrate rational processing with Spark's functional programming API and MLlib \cite{meng2016mllib} to work with structured and semistructured data using either SQL or DataFrame API. In Spark, DataFrame is a distributed column-based collection of data. In comparison with a table in a relational database or a data frame in R/Python, it is similar to both conceptually, but with richer optimizations under the hood. DataFrames can be constructed from a wide array of sources such as structured data files, tables in Hive, external databases, or existing RDDs. Additionally, Spark has a catalyst layer to optimize the execution plan of SQL queries. 
	\item
	\textit{Interactive processing} is a system that gives users a way to undertake their own analysis in an interactive manner \cite{chen2012interactive}. In an interactive analytical processing framework, users can interact with systems directly, and review, compare, and analyze the input data in tabular or graphic format. Google's Dremel \cite{melnik2010dremel}, Apache drill \cite{hausenblas2013apache} and Apache Spark \cite{zaharia2010spark} are distributed systems for interactive analysis of data-intensive computing. One of Spark's most compelling features is its capability for interactive analytics. Through this feature, developers can incorporate a variety of Spark libraries, such as Spark Streaming for visualizing streaming, machine learning algorithms \cite{meng2016mllib} for iterative tasks, and GraphX \cite{xin2013graphx} for displaying graph analyses.
	
\end{enumerate}

\section{Performance Defects}
\label{sec:performance}

The availability of massive volume of data has led to the proliferation of data-intensive applications. The mainstream approach to scalability and expandability is to distribute data and computation to a large number of machines so that multiple processing nodes can work simultaneously. In order to process growing datasets efficiently, there exists a large body of techniques \cite{diersen2011classification,fang2015interruptible,guo2011model,guo2015accurate,huang2013scalable,huang2012mpi, ma2013symbolic,mitchell2007causes, xu2012finding, xu2010finding, zhang2017mrapid} that spans a variety wide of disciplines to improve data-intensive computing system performance. In this paper, we survey promising semantics-aware approaches to optimize the performance from the perspective of program analysis. 

\subsection{Data Access}
With the dramatic increase of data size, scientists and engineers take great efforts to deal with data-intensive issues. Apart from considerable computational needs, tremendous I/O operations are also required. There are several ways to improve I/O performance. One of the promising methods is to reduce disk access during execution. This can be achieved by two approaches: 1) caching the frequently used data in memory instead of disks; 2) restructuring application code in a way that maximizes data reuse. 
An approach \cite{kandemir2008improving} to improve I/O performance is by reducing disk access through a new concept called disk reuse maximization. In this compiler-based approach, it uses a polyhedral tool to analyze data dependencies in the application code to maximize data reuse in a given set of disks as much as possible before moving them to other disks. Another approach \cite{chen2014decoupled} shows a new way of moving computations near to data in order to minimize data movement by decoupling I/O to address the I/O bottleneck issues via using compiler technologies.

\subsection{Memory Management}

In order to speed up development cycle and provide a friendly application interface, most data-intensive systems are developed in managed languages, such as C\# and Java. Even though, there is an automatic memory management for some programming languages. Memory management in a data-intensive system is often prohibitively expensive. For instances, allocating and de-allocating a set of data objects would consume a huge of memory, which leads to poor performance of runtime system. In this scenario, systems could incur a high memory management overhead to allocate and release memory, and prolong the execution time. The computation on a worker node often suffers extensive memory pressure, \ie, the heap's limit is reached and more memory is required. Data-intensive applications may crash because of out-of-memory errors. The execution time can also be affected by garbage collection (GC), which is another challenging issue for performance optimization.  In order to fix out or alleviate memory pressure, several memory optimization approaches have been proposed. FACADE \cite{nguyen2015facade} is a novel compiler framework, which tries to automatically transform data paths of an existing data-intensive application to generate highly-efficient data manipulation code. In FACADE, the number of runtime heap objects created for data types in each thread is (almost) statically bounded, which reduces memory management cost and improves scalability. In \cite{nguyen2016yak}, a garbage collector called Yak is designed to provide high throughput and low latency for all JVM-based languages. In Yak, the management heap is divided into a control space (CS) and a data space (DS) based on the observation that there is a clear distinction between a control path and a data path in a typical data-intensive system.

\subsection{Data Shuffle}

In oder to remedy the drawback of CPU clock frequency,  computations and data can be distributed on a larger number of commodity computers to improve the performance of data-intensive applications \cite{dean2008mapreduce,zhang2014smarth}. In such data parallel programs, data shuffling among computers can dominate the whole program performance. In recent years, how to reduce data shuffling is an active research area \cite{davidson2013optimizing,guo2012spotting,zhang2012optimizing}. In \cite{zhang2012optimizing}, a few useful properties for User Defined Function (UDF) are identified to reason about data-partition properties across phases. \cite{guo2012spotting} proposes a series of semantics-aware optimizations on data-intensive program's procedural code, such as data filtering, eliminating unnecessary code and data and calculating small derived values earlier, to minimize the amount of data-shuffling between the pipeline stages of a distributed data parallel program. 

\subsection{Data Analysis}
In data-intensive computing, it is crucial to generate an efficient execution plan based on properties of code, data and platform. It may lead to a poor performance by using a fixed priori experience about these properties to determine execution plans, as most current platforms do. Moreover, it is difficult to extract and estimate these properties according to the highly distributed nature of data-intensive computing frameworks and the freedom that allows users to use UDF to represent a series of data operations. In \cite{agarwal2012re}, a framework, namely RoPE (Reoptimizer for Parallel Executions), is proposed to collect code and data properties by piggybacking on job execution. Then it determines execution plans by feeding these properties to a query optimizer component. In \cite{jahani2011automatic}, a framework called MANIMAl automatically analyzes MapReduce programs and applies appropriate data-aware optimizations to the programs. 

\section{Research Challenges and Opportunities}
\label{sec:research}

Data-intensive computing is playing a critical role in transforming economies and delivering a new wave of productive growth \cite{manyika2011big}. 
While data-intensive computing brings many attractive benefits, it is also facing grand challenges \cite{ahrens2011data,chen2014data, katal2013big,khan2014big, labrinidis2012challenges,michael2013big} and research opportunities. To tackle data-intensive computing problems, most difficulties lie in data capture, storage, searching, sharing, analysis, and visualization \cite{chen2014big}. We classify these challenges into three categories: data management, data analytics, and infrastructure issues.


\subsection{Data Management}

Data management confronts many issues about massive amount of heterogeneous and complex data. 
\begin{enumerate}
	\item{\textit{Data Representation:}}
    Because of diverse data sources, datasets often include certain levels of heterogeneity such as type, structure, semantics, organization, granularity, and accessibility. The target of data representation is to make data more meaningful for computer analysis and user interpretation. It is inefficient for users to do analytics from an improper data representation since it may reduce the value of original data. Therefore, there should be a competent data representation to reflect structure, hierarchy, and variety of the data 
to enable efficient operations on different datasets.

	\item{\textit{Data Reduction and Compression:}}
	How to remove redundant data in raw datasets and compress them without losing potential value is critical to reduce  overhead and improve overall system performance. Due to the enormous size of raw datasets, it is necessary to reduce this huge volume of data into a manageable size for a storage system. In addition, it is also necessary to remove duplicated data for processing data efficiently. Although there are already data reduction methods such as dimension reductions techniques to reduce data size, there are many research opportunities on redundancy elimination and compression-based reduction.
    
\item{\textit{Data Life-Cycle Management:}} The amount of digital data increases at an unprecedented rate, so one of the urgent challenges is that there is no suitable storage system to support accessing the huge size of data in an efficient way. Generally speaking, the value behind data depends on data freshness, hence it is critical to design algorithms to decide which data shall be kept and which data shall be discarded. Besides that, a novel storage system is needed to support accessing, searching, moving, and sharing data in an efficient and scalable way.
\end{enumerate}

\subsection{Data Analytics}
\begin{enumerate}
	\item{\textit{Expandability and Scalability:}}
	The first impression of data-intensive computing is the massive size of data. Therefore, the most important challenge is how to scale up analytical algorithms to process more complex datasets and scale out horizontally to support increasingly expanding datasets.
	\item{\textit{Timeliness:}}
	For those real-time data-intensive applications, like navigation, social networks, Internet of Thing, it is critical for data-intensive computing system to ensure the timeliness of response when the volume of data to be processed is very large. Search is a frequent operation to find elements that meet a specified criterion. 
The complexity of time and space for the search algorithm on massive volume of data is a challenging issue, especially for real-time systems. 
	\item{\textit{Data Privacy and Security:}}
	With the proliferation of online and mobile services, privacy and security concerns are emerging regarding accessing and analyzing personal information. 
It is important to enhance systems to eliminate privacy leakage and security issues without impeding analyses and affecting system performance. This requires a comprehensive solution including network, software system, and data.

\end{enumerate}

\subsection{Infrastructure Issues}
\begin{enumerate}
	\item{\textit{Computer Architecture:}}
    In recent years, scientific computing turns to use co-processors (accelerators) to combat limitation of CPU. For example, graphics processing units (GPUs) has been widely used to speed up numerical computation in many areas such as scientific modeling and machine learning. However, it is not easy to implement applications and gain ideal performance speedup on GPU. On the lower level of programming models, CUDA provides an efficient way to enable performance acceleration on Nvidia GPUs. On the higher level of programming platforms, Tensorflow is one of the most popular libraries to enable machine learning to utilize powerful GPUs. To compare with GPUs, Intel Xeon Phi Many Integrated Core (MIC) is a homogeneous architecture with more cores and hardware threading than in a regular processor. A key advantage of MIC is to support general parallel programming models and languages, such as OpenMP and pThread, to run on many cores.

	\item{\textit{Storage System:}}
	How to efficiently store and access massive volume of data is still a critical issue for industry and academia. For the existing storage architectures, there are some severe drawbacks and limitations to support data-intensive computing systems very well. For example, even though the development of solid-state drive (SSD) and phase-change memory (PCM) helps mitigate the difficulties, these newer technologies fail to deliver equal speed for random and sequential I/O access, 
    which leads to think over how to design a novel layered storage subsystem for data-intensive computing systems.
    
	\item{\textit{Data Transmission:}}
    The key point to improve the performance of data-intensive computing system is how to minimize data shuffling between two different stages, which depends on the network bandwidth and data volume transmitted between stages. From this point of view, network capacity is one of bottlenecks in data-intensive computing systems, especially when the volume of communication is heavy. Considering the network bottlenecks, researchers have been proposing a great deal of approaches to improve the efficiency of data transmission, such as restructuring application code in a way to maximize data reuse, optimizing the execution plans to filter or remove unused data before shuffling operations. But it is still a challenging issue to understand the semantics of application code and data, which may facilitate potential optimizations on software-defined environment.
	\item{\textit{Energy Management:}}
	As data volume and analytics demand expand dramatically, there is an increasing consumption on energy. Especially in IoT devices with limited energy supply, this issue is becoming more critical. Such a kind of energy consumption in large scale computing systems has attracted growing attentions. Therefore, a system-level power control and management mechanism is needed for data-intensive computing systems without affecting their extensibility and accessibility.
	\item{\textit{Multi-Discipline:}}
	It is apparent that data-intensive analytics is an interdisciplinary field that requires expertise from different domains to collaborate to mine hidden values. Hence, a sophisticated cooperation among various disciplines is needed to explore technologies in data analytics. Scientific workflow management systems provide a good way to connect multiple domain experts in the support of composing and executing a series of computational or data manipulation steps in a scientific application.
    
\end{enumerate}
\section{Conclusions}
\label{sec:conclusion}
In order to tackle massive volume of data, a wide variety of data-intensive computing systems have been proposed. Regarding the performance optimization of these platforms, scientists and researchers from various disciplines have developed a large body of techniques to improve overall performance of systems. However, there are still performance defects among these systems. In order to provide a comprehensive understanding of data-intensive computing, especially semantics-aware methods to improve system performance, in this study we give a thorough overview including common characteristics and paradigm shifts in th evolution, the promising programming models and technologies, a classification of performance defects from the perspective of semantics-aware approaches, as well as research challenges and opportunities. While the approaches mentioned in Section \ref{sec:performance} have been demonstrated to be effective, there is still much room to explore and improve system performance. For example, a major issue is network communication and data shuffling. In our future research, we will investigate a systematic approach based on semantics-aware technology to reduce communication between nodes and minimize the amount of data shuffling between pipelined stages. In addition, we will enhance execution plans and conduct in-situ optimization based on semantics extracted from data, code, and system profiling. 

\section*{Acknowledgment}
\label{sec:akw}
This work was supported in part by NSF-CAREER-1622292.

\begin{spacing}{0.45}
\bibliographystyle{abbrv}
\bibliography{reference}
\end{spacing}

\end{document}